\documentclass[journal,twocolumn]{IEEEtran}
\usepackage{amssymb}
\usepackage{amsmath}
\usepackage{amsthm}
\usepackage{amsfonts}
\usepackage{graphicx}
\usepackage[numbers, square, comma, sort&compress]{natbib}
\usepackage{xcolor}
\usepackage[nolist]{acronym}
\usepackage[bookmarks=false]{hyperref}
\usepackage{caption}
\usepackage{subcaption}
\usepackage{comment}
\usepackage{csquotes}
\usepackage{tikz}
\usepackage{bm}
\usetikzlibrary{arrows,shapes.misc,chains,scopes}
\usepackage{pgfplotstable}
\pgfplotsset{compat=1.15}
\usepackage[ruled,linesnumbered,vlined]{algorithm2e}
\SetNlSty{bfseries}{\color{black}}{}
\SetArgSty{textnormal}
\SetKw{Continue}{continue}
\usepackage{enumitem}
\setlist[itemize]{leftmargin=*}
\SetKw{Break}{break}

\newcommand{\noiseguess}[1]{z^{n,{#1}}}
\newcommand{\rvnoise}{N^n}
\newcommand{\noise}{z^n}
\newcommand{\rvchanout}{R^n}
\newcommand{\rvchanouti}{R_i}
\newcommand{\chanout}{r^n}
\newcommand{\chanouti}{r_i}
\newcommand{\rvdemodout}{Y^n}
\newcommand{\demodout}{y^n}
\newcommand{\codebook}{\mathcal{C}}
\newcommand{\codeword}{c^n}
\newcommand{\codewordi}{c_i}
\newcommand{\cleaned}{\hat{x}^n}
\newcommand{\rvword}{C^n}

\newcommand{\binaryAlphabetN}{\{0,1\}^n}

\newcommand{\mlcodeword}{c^{n,*}}

\newcommand{\decodelist}{\mathcal{L}}

\graphicspath{{./img/}}

\title{Soft-output Guessing Codeword Decoding}

\begin{document}

\author{Ken~R.~Duffy, 
Peihong~Yuan, 
Joseph~Griffin,
       and~Muriel M\'edard
\thanks{P. Yuan, J. Griffin, and M. M{\'e}dard, Massachusetts Institute of Technology (e-mails: \{phyuan, joecg, medard\}@mit.edu).}
\thanks{K. R. Duffy, Northeastern University, (e-mail: k.duffy@northeastern.edu).}
\thanks{This work was supported by the Defense Advanced Research Projects Agency (DARPA) under Grant HR00112120008.}

}

\begin{acronym}
    \acro{PC}{product code}
    \acro{BCH}{Bose-Chaudhuri-Hocquenghem}
    \acro{eBCH}{extended Bose-Chaudhuri-Hocquenghem}
    \acro{GRAND}{guessing random additive noise decoding}
    \acro{SGRAND}{soft GRAND}
    \acro{DSGRAND}{discretized soft GRAND}
    \acro{SRGRAND}{symbol reliability GRAND}
    \acro{ORBGRAND}{ordered reliability bits GRAND}
    \acro{SOGRAND}{soft-output GRAND}
    \acro{5G}{the $5$-th generation wireless system}
    \acro{NR}{new radio}
	\acro{APP}{a-posteriori probability}
	\acro{ARQ}{automated repeat request}
	\acro{ASK}{amplitude-shift keying}
	\acro{AWGN}{additive white Gaussian noise}
	\acro{B-DMC}{binary-input discrete memoryless channel}
	\acro{BEC}{binary erasure channel}
	\acro{BER}{bit error rate}
	\acro{biAWGN}{binary-input additive white Gaussian noise}
	\acro{BLER}{block error rate}
	\acro{uBLER}{undetected block error rate}
	\acro{bpcu}{bits per channel use}
	\acro{BPSK}{binary phase-shift keying}
	\acro{BSC}{binary symmetric channel}
	\acro{BSS}{binary symmetric source}
	\acro{CDF}{cumulative distribution function}
	\acro{CRC}{cyclic redundancy check}
	\acro{DE}{density evolution}
	\acro{DMC}{discrete memoryless channel}
	\acro{DMS}{discrete memoryless source}
	\acro{BMS}{binary input memoryless symmetric}
	\acro{eMBB}{enhanced mobile broadband}
	\acro{FER}{frame error rate}
	\acro{uFER}{undetected frame error rate}
	\acro{FHT}{fast Hadamard transform}
	\acro{GF}{Galois field}
	\acro{HARQ}{hybrid automated repeat request}
	\acro{i.i.d.}{independent and identically distributed}
	\acro{LDPC}{low-density parity-check}
	\acro{GLDPC}{generalized low-density parity-check}
	\acro{LHS}{left hand side}
	\acro{LLR}{log-likelihood ratio}
	\acro{MAP}{maximum-a-posteriori}
	\acro{MC}{Monte Carlo}
	\acro{ML}{maximum-likelihood}
	\acro{PDF}{probability density function}
	\acro{PMF}{probability mass function}
	\acro{QAM}{quadrature amplitude modulation}
	\acro{QPSK}{quadrature phase-shift keying}
	\acro{RCU}{random-coding union}
	\acro{RHS}{right hand side}
	\acro{RM}{Reed-Muller}
	\acro{RV}{random variable}
	\acro{RS}{Reed–Solomon}
	\acro{SCL}{successive cancellation list}
	\acro{SE}{spectral efficiency}
	\acro{SNR}{signal-to-noise ratio}
	\acro{UB}{union bound}
	\acro{BP}{belief propagation}
	\acro{NR}{new radio}
	\acro{CA-SCL}{CRC-assisted successive cancellation list}
	\acro{DP}{dynamic programming}
	\acro{URLLC}{ultra-reliable low-latency communication}
    \acro{GCC}{Generalized Concatenated Code}
    \acro{GCCs}{Generalized Concatenated Codes}
    \acro{MDS}{maximum distance separable}
    \acro{ORBGRAND-AI}{ORBGRAND Approximate Independence}
    \acro{BLER}{block error rate}
    \acro{crc}[CRC]{cyclic redundancy check}
    \acro{cascl}[CA-SCL]{CRC-Assisted Successive Cancellation List}
    \acro{uer}[UER]{undetected error rate}
    \acro{MDR}{misdetection rate}
    \acro{LMDR}{list misdetection rate}
    \acro{QC}{quasi-cyclic}
    \acro{SISO}{soft-input soft-output}
    \acro{VN}{variable node}
    \acro{CN}{check node}
    \acro{SI}{soft-input}
    \acro{SO}{soft-output}
    \acro{PAC}{polarization-adjusted convolutional}
    \acro{dRM}{dynamic Reed Muller}
    \acro{QPSK}{quadrature phase shift keying}
\end{acronym}
\maketitle

\begin{abstract}

We establish that it is possible to extract accurate blockwise and bitwise soft output from Guessing Codeword Decoding with minimal additional computational complexity by considering it as a variant of Guessing Random Additive Noise Decoding. Blockwise soft output can be used to control decoding misdetection rate while bitwise soft output results in a soft-input soft-output decoder that can be used for efficient iterative decoding of long, high redundancy codes.
\end{abstract}

\begin{IEEEkeywords}
GCD, GRAND, Soft Output, SISO, concatenated codes.
\end{IEEEkeywords}

\section{Introduction}
For any soft input (SI) forward error correction decoder, it would be highly desirable if it could produce accurate blockwise and bitwise soft output (SO). Blockwise SO, in the form of an a posteriori likelihood that the decoding is correct, can be used to inform retransmission requests or to label untrustworthy blocks for an erasure correction code to rectify \cite{lin_error_2004}, while bitwise SO can be used for efficient, iterative soft input soft output (SISO) decoding of long, high redundancy codes such as product codes~\cite{elias_error-free_1954,pyndiah_1998}, \ac{LDPC} codes \cite{gallagher1962_lowdensity, costello2007channel}, and \ac{GLDPC} codes~\cite{liva2008quasi,Lentmaier10}. Most error correction decoding algorithms, however, cannot generate accurate SO for reasons articulated in seminal work of Forney \cite{forney1968_exponential}.

Consider a codebook, $\codebook\subset\{0,1\}^n$, consisting of $2^k$ binary codewords of length $n$, where transmitted codewords are selected uniformly at random. Given continuous-valued channel output $\chanout$, which serves as SI for the decoder, and a codeword, $\mlcodeword\in\codebook$, the posterior likelihood that $\mlcodeword$ was the transmitted can be expressed as
\begin{align}
    p_{\rvword|\rvchanout}(\mlcodeword|\chanout) = \frac{f_{\rvchanout|\rvword}(\chanout|\mlcodeword)}{\displaystyle\sum_{\codeword\in\codebook} f_{\rvchanout|\rvword}(\chanout|\codeword)}. \label{eq:trueSO}
\end{align}
The denominator is a sum consisting of $2^k$ terms whose evaluation is impractical for even moderate $k$, which is the central problem. Given a list, $\decodelist \subseteq \codebook$, of the $L\geq 2$ most likely codewords, in Forney's study he considered the approximation 
\begin{align}
&p_{\rvword|\rvchanout}(\mlcodeword|\chanout) \nonumber\\
&\approx \frac{f_{\rvchanout|\rvword}(\chanout|\mlcodeword)}{\displaystyle\sum_{\codeword\in\decodelist} f_{\rvchanout|\rvword}(\chanout|\codeword)
+ \underbrace{\sum_{\codeword\in\codebook\setminus\decodelist}f_{\rvchanout|\rvword}(\chanout|\codeword)}_{\text{assumed to be } 0}}, \label{eq:missingterm}
\end{align}
which also underlies the approximate bitwise SO introduced by Pyndiah for iterative decoding of product codes \cite{pyndiah_1998}. Guessing Codeword Decoding (GCD) \cite{ma2024guessing,zheng2024universal}, which can decode any binary linear code, generates a list of candidate codewords and their likelihoods as part of its execution and so it may be tempting to use Forney's approximation to generate SO from it, but we shall demonstrate that inaccurate SO is produced.

In the context of SI Guessing Random Additive Noise Decoding (GRAND) algorithms, e.g. \cite{Duffy19,duffy22ORBGRAND,Duffy23ORBGRANDAI,Liuetal23}, we have recently developed a method for extracting accurate blockwise and bitwise SO that obviates the need for a decoding list and can, distinctively, provide SO with a single decoding~\cite{galligan2023upgrade,yuan2023soft}. Central to the approach is the development of a natural estimate of the
term in eq. \eqref{eq:missingterm} that is assumed to be $0$ in the Forney and Pyndiah approximations. This estimate is dynamically evaluated during the decoding process with minimal additional computation.  Building on the approach, we have since established that accurate SO can also be produced by all successive cancellation (SC)-based decoders of polar-like codes \cite{yuan2024nearoptimal}. 

By considering Guessing Codeword Decoding (GCD) \cite{ma2024guessing,zheng2024universal} as a variant of GRAND, here we establish that it is possible to extract accurate blockwise and bitwise SO from it using minimal additional computation. In section \ref{sec:SOGRAND} we introduce SI GRAND algorithms and how SO is extracted from them. In \ref{sec:SOGCD} we show how SO can be generated during GCD by considering it as a variant of GRAND. Section \ref{sec:SO} establishes the accuracy of the blockwise and bitwise SO, while Section \ref{sec:peva} provide SISO performance results.

\section{GRAND and SO}
\label{sec:SOGRAND}

Let $\codebook$ be a codebook containing $2^k$ binary codewords each of length $n$ bits. Let $\rvword$ be a codeword drawn uniformly at random from the codebook and let $\rvnoise$ denote the binary noise effect that the channel has on that codeword during transmission. I.e., $N^n$ encodes the difference between the demodulated received sequence and the transmitted codeword. Then, $\rvdemodout = \rvword \oplus \rvnoise$ is the demodulated channel output, with $\oplus$ being the element-wise binary addition operator. Let $\rvchanout$ denote real soft channel output per-bit.
Lowercase letters represent realizations of random variables, with the exception of $\noise$, which is the realization of $\rvnoise$, which is assumed independent of the channel input.

GRAND algorithms operate by progressing through a series of noise effect guesses $\noiseguess{1}, \noiseguess{2}, \ldots \in \binaryAlphabetN$, whose order is informed by channel statistics or SI, e.g. \cite{Duffy19a, An21, An22, duffy2022_ordered,Duffy23ORBGRANDAI}, until it finds one or more, $\noiseguess{q}$, that satisfy $\cleaned_q = \demodout \ominus \noiseguess{q} \in \codebook$, where $\ominus$ inverts the effect of the noise on the channel output. If the guesses are in decreasing order of likelihood, then the the $l$-th such $\noiseguess{q}$ is the $l$-th most likely estimate of $\rvnoise$ and $\cleaned_q$ is the $l$-th most likely estimate of the transmitted codeword $\rvword$ \cite{abbas2021list,condo2022_iterative,galligan2023_block}. Note that, in contrast to, e.g., successive cancellation list decoders or GCD, this approach is a true list decoder in the sense that, should the query order go from most likely to least likely, the list sequentially contains the most likely codewords.

As the guessing procedure does not depend on codebook structure, GRAND can decode any moderate redundancy code of any structure, including non-linear ones, as long as it has a method for checking codebook membership. For example, for a linear block code with an $(n-k)\times n$ parity-check matrix $\mathbf{H}$, $\cleaned_q$ is a codeword if $\mathbf{H} \cleaned_q = 0^n$ \cite{lin_error_2004}, where $\cleaned_q$ is taken to be a column vector and $0^n$ is the zero vector. 

Underlying GRAND is a race between the number of guesses until the true codeword would be identified and the number of guesses until an incorrect codeword would be identified. Whichever of these finishes first determines whether the decoding identified by GRAND is correct. By noting that Eq. \eqref{eq:trueSO} can be rewritten as
\begin{align}
    p_{\rvword|\rvchanout}(\mlcodeword|\chanout) 
    & = \frac{p_{N^n|\rvchanout}\left(\mlcodeword \oplus \demodout \middle| \chanout\right)}
    {\displaystyle\sum_{\codeword\in\codebook} p_{N^n|\rvchanout}\left(\codeword \oplus \demodout \middle| \chanout\right)} \label{eq:trueSO2}
\end{align}
and analysing the race between those two random variables, the following SO for GRAND is developed and then applied in \cite{galligan2023upgrade,yuan2023soft}. 

For any noise effect query order, regardless of whether it is from most likely to least likely, assume that each of $L$ codewords are identified at query numbers $q_1,q_2,\ldots,q_L$, so that $c^{n,i} = \demodout\oplus z^{n,q_i}$, then the likelihood that $c^{n,i}$ is the correct decoding, i.e. eq. \eqref{eq:trueSO}, can be approximated as
\begin{align}
\frac{\displaystyle p_{N^n|\rvchanout}(z^{n,q_i}|\chanout) }
{
\begin{array}{l}
\displaystyle 
\sum_{i=1}^Lp_{N^n|\rvchanout}(z^{n,q_i}|\chanout) \\
\displaystyle \qquad + \left(1-\sum_{j=1}^{q_L} p_{N^n|\rvchanout}(z^{n,j}|\chanout)\right) \left(\frac{2^k-1}{2^n-1}\right) 
\end{array}
}.
\label{eq:SOGRAND}
\end{align}
The final term in the denominator essentially replaces the term corresponding to the one set to zero in eq. \eqref{eq:missingterm}. 

To evaluate the SO in eq. \eqref{eq:SOGRAND} it suffices to record: a running sum of the likelihood of all noise effect queries that have been made; the likelihoods of the noise effect queries that result in codewords being identified; and the code's dimensions, $[n,k]$. In~\cite{galligan2023upgrade,yuan2023soft} it is shown that this blockwise SO approximation, and the bitwise version that results, is accurate for codes of any structure and can be used in a variety of powerful applications.

\section{GCD and SO}
\label{sec:SOGCD}
While GRAND algorithms can decode any code of any structure, including non-linear ones created with length constraints and  cryptographic functions \cite{ozaydin2022grand,cohen2023aes,woo2024leveraging}, proposals have been made to exploit the structure of binary linear codes to influence the query order, e.g. \cite{rowshan2022_constrained,rowshan2023lowcomplexity}. These approaches result in identical decoding accuracy with fewer noise effect queries. For example, for even linear block codes~\cite{lin_error_2004}, which includes \ac{eBCH} codes, polar codes \cite{arikan2009} and many others, the parity of the noise effect pattern being sought is the same as the parity of the hard demodulated signal. Armed with a noise effect generator that can skip given Hamming weights, such as the landslide algorithm in Ordered Reliability Bits GRAND (ORBGRAND) \cite{duffy2022_ordered,Riaz23}, that property can be used to reduce the total query number by a factor of up to $2$ while providing identical decoding performance as if the missing queries were made.

While not originally described in those terms, GCD \cite{ma2024guessing,zheng2024universal} can be thought of as a highly effective way to use the structure of the codebook to make $2^{n-k}$ noise effect guesses at a time, where one of the noise effect guesses will necessarily result in a codeword and $2^{n-k}-1$ of them will not. If a GRAND algorithm does not avail of codebook structure to reduce query numbers for a binary linear code, it is proven in \cite{ma2024guessing} that the maximum likelihood decoding is identified in fewer queries by GCD. When, for example, decoding an even binary linear code with a GRAND variant that avails of that structure, however, that is not necessarily the case for codes with moderate $n-k$. 

GCD is most readily understood for systematic binary linear codes where the first $k$ bits are the information bits. Noise effects are sequentially queried from most likely to least likely on the systematic bits, $z^{k,1},z^{k,2},\ldots$ and then extended via the generator matrix, $G$, to create a codeword $(y^k\oplus z^{k,i})G = c^{n,i}$. No guarantees are made about the likelihoods of the individual $c^{n,i}$, but the key ingenuity is the observation of a termination condition where the sequence of noise effect queries can cease when 
\begin{align}
p_{N^k|R^k}(z^{k,i+1}|r^k)\leq \max_{j\in\{1,\ldots,i\}} p_{\rvword|\rvchanout}(c^{n,j}|\chanout)
\label{eq:GCDtermination}
\end{align}
as any further query will necessarily result in a less likely codeword than the most likely one that has already been identified. If the code is non-systematic, a single advance operation that identifies a linear transform suffices to make the process equivalent to the systematic one. In this way, GCD circumvents the two key practical challenges of Ordered Statistics Decoding by: not requiring Gaussian Elimination, an $O(n^3)$ procedure, with every decoding; and having a readily tested  termination condition for decoding.

Viewed through a GRAND lens, however, each noise effect query on $k$ bits effectively results in $2^{n-k}$ queries, those corresponding to all $2^{n-k}$ possible extensions of which only one can result in a codeword. Notably, as the GRAND SO formula in eq. \eqref{eq:SOGRAND} holds for an arbitrary query order, it applies when GCD is considered as a GRAND variant.
The combined likelihood of all the length $n$ extrapolated noise effect queries made from a noise effect applied to the $k$ bits, $z^{k,i+1}$, is already evaluated for the left hand side of the termination condition eq. \eqref{eq:GCDtermination}.  Explicitly, when considering GCD as a GRAND variant, the likelihood that the codeword identified on the $j$-th query of $L$ queries is correct is, approximately,
\begin{align}
&p_{\rvword|\rvchanout}(c^{n,j}|\chanout) \nonumber\\
&\approx
\frac{\displaystyle p_{N^n|\rvchanout}(c^{n,j}\oplus \demodout|\chanout) }
{
\begin{array}{l}
\displaystyle 
\sum_{i=1}^Lp_{N^n|\rvchanout}(c^{n,i}\oplus \demodout|\chanout) \\
\displaystyle \qquad + \left(1-\sum_{j=1}^{L} p_{N^k|R^k}(z^k|r^k)\right) \left(\frac{2^k-1}{2^n-1}\right) 
\end{array}
}.
\label{eq:SOGCD}
\end{align}
The likelihood that the decoding in not in the list of identified codewords is then simply 
\begin{align}
    p_{\rvword|\rvchanout}\left(\codebook{\setminus}\decodelist|\chanout\right) = 
    1-\sum_{i=1}^L p_{\rvword|\rvchanout}(c^{n,j}|\chanout).
    \label{eq:SOGCD2}
\end{align}

In order to evaluate eq. \eqref{eq:SOGCD} and eq. \eqref{eq:SOGCD2}, the same information is required as in  eq. \eqref{eq:SOGRAND}, which is already available to the decoder: the likelihood of each codeword that is identified; the sum of the likelihoods of all noise effect sequences that have been queried, including those that do not lead to a codeword; and the code's dimensions.

Based on the blockwise SO in eq. \eqref{eq:SOGCD} and eq. \eqref{eq:SOGCD2}, the bitwise SO LLR for the $i$-th bit is
\begin{align}
&\log 
    \frac{\displaystyle
    \sum_{\codeword\in\decodelist:\codewordi=0} p_{\rvword|\rvchanout}(\codeword|\chanout) 
        + p_{\rvword|\rvchanout}\left(\codebook{\setminus}\decodelist|\chanout\right) p_{X|\rvchanouti}(0|\chanouti)
     }
     {\displaystyle 
     \sum_{c^n\in\decodelist:\codewordi=1} p_{\rvword|\rvchanout}(\codeword|\chanout) + p_{\rvword|\rvchanout}\left(\codebook{\setminus}\decodelist|\chanout\right) p_{X|\rvchanouti}(1|\chanouti)
     } \label{eq:LLRi}
\end{align}
which reflects the content of each codeword in the list and its likelihood, in addition to the likelihood that the codeword is not found, whereupon the prior information is retained.

While GCD terminates after finding the most likely decoding, it generates a list in the process. That list, however, need not contain other likely decodings. GCD's termination condition can be changed to stop after a number of highly likely codewords have been found by replacing the one in eq. \eqref{eq:GCDtermination} to one where, if a list of $\Lambda$ highly likely (and hence good, from a list-decoding point of view) codewords is needed, the right-hand side corresponds to the $\Lambda$-th most likely codeword in the list \cite{zheng2024universal}. In what follows, we demonstrate that this blockwise SO in eq. \eqref{eq:SOGCD} is highly accurate for Random Linear Codes (RLCs), and for highly structured extended BCH (eBCH) codes once $\Lambda$ is larger than $1$. The latter is necessary to create useful bitwise SO for iterative SISO decoding applications.

\section{Noise effect sequences}

As with SGRAND \cite{solomon20}, GCD provides optimally accurate maximum likelihood decoding when noise effect queries are sequentially created from most likely to least likely, as in algorithms described in e.g. \cite{solomon20,ma2024guessing,zheng2024universal}. Algorithms for generating those noise effect sequences, however, require involved dynamic memory for processing, for which no efficient circuits have been published. ORBGRAND \cite{Duffy21_ordered,duffy22ORBGRAND,abbas2021orbgrand,Liuetal23} produces noise effect sequences in approximately the decreasing order of likelihood with no dynamic memory, making it suitable for highly efficient implementation in hardware \cite{Riaz23}. To distinguish GCD driven by ORBGRAND's 1-line order, we refer to it as GCD-ORB.

\section{SO-GCD accuracy}
\label{sec:SO}

\begin{figure}
\centering
	\footnotesize
	\begin{tikzpicture}[scale=1]
\footnotesize
\begin{loglogaxis}[
legend style={at={(0.5,0.5)},anchor= north west},
ymin=1e-4,
ymax=1,
width=3.5in,
height=3.5in,
grid=both,
xmin = 1e-4,
xmax = 1,
xlabel = {predicted},
ylabel = {empirical},
]

\addplot[blue,mark = +,dashed,mark options=solid]
table[]{x y
0.57819 0.56258
0.19204 0.17702
0.062769 0.057353
0.0203 0.018287
0.0068922 0.0054388
0.0023454 0.0020139
};\addlegendentry{$(64,57)$, $\Lambda=1$}

\addplot[red,mark=o,dashed,mark options=solid]
table[]{x y
0.50772 0.51546
0.179 0.18344
0.060037 0.053456
0.019387 0.016356
0.006511 0.0058002
0.0022598 0.0020866
0.0008192 0.0006887
};\addlegendentry{$(32,26)$, $\Lambda=1$}

\addplot[brown,mark=x,dashed,mark options=solid]
table[]{x y
0.4881 0.4729
0.17118 0.14647
0.059308 0.045497
0.018931 0.014465
0.0061901 0.0048663
0.002081 0.0015891
0.00070105 0.00053248
};\addlegendentry{$(16,11)$, $\Lambda=1$}

\addplot[blue,mark = +]
table[]{x y
0.57582 0.56751
0.19534 0.20788
0.061032 0.062225
0.018976 0.017056
0.0063226 0.0063732
};\addlegendentry{$(64,57)$, $\Lambda=2$}

\addplot[red,mark=o]
table[]{x y
0.49672 0.50682
0.1903 0.17997
0.058688 0.05
0.018858 0.017785
0.0060543 0.0053197
0.0019386 0.00153
};\addlegendentry{$(32,26)$, $\Lambda=2$}

\addplot[brown,mark=x]
table[]{x y
0.45976 0.46895
0.18968 0.19546
0.058928 0.059496
0.018713 0.019912
0.0060045 0.0063359
0.0019205 0.0020758
0.00061792 0.00062286
0.00019944 0.00017251
};\addlegendentry{$(16,11)$, $\Lambda=2$}

\addplot[blue,mark = +,dashed,mark options=solid]
table[]{x y
0.20585 0.3908
0.10461 0.29721
0.032149 0.14867
0.010245 0.066548
0.0032436 0.031429
0.0010409 0.013763
0.0003266 0.0040816
};\addlegendentry{$(64,57)$, Forney, $\Lambda=2$}

\addplot[red,mark=o,dashed,mark options=solid]
table[]{x y
0.20428 0.31595
0.099065 0.20113
0.030675 0.085193
0.0096683 0.033944
0.0031101 0.014336
0.00099809 0.0059717
0.00031773 0.002382
};\addlegendentry{$(32,26)$, Forney, $\Lambda=2$}

\addplot[brown,mark=x,dashed,mark options=solid]
table[]{x y
0.20356 0.25253
0.097707 0.13999
0.030605 0.054924
0.0098509 0.0214
0.003179 0.009148
0.0010311 0.0045293
0.00032468 0.0022046
};\addlegendentry{$(16,11)$, Forney, $\Lambda=2$}

\addplot[gray,dashed]
table[]{x y
1e-5 1e-5
1 1
};

\end{loglogaxis}

\end{tikzpicture}
	\caption{SO Predicted BLER vs. empirical BLER: RLCs, SO-GCD-ORB, $E_b/N_0=3$.}
	\label{fig:blkSO_acc_listGCD_RLC}
\end{figure}
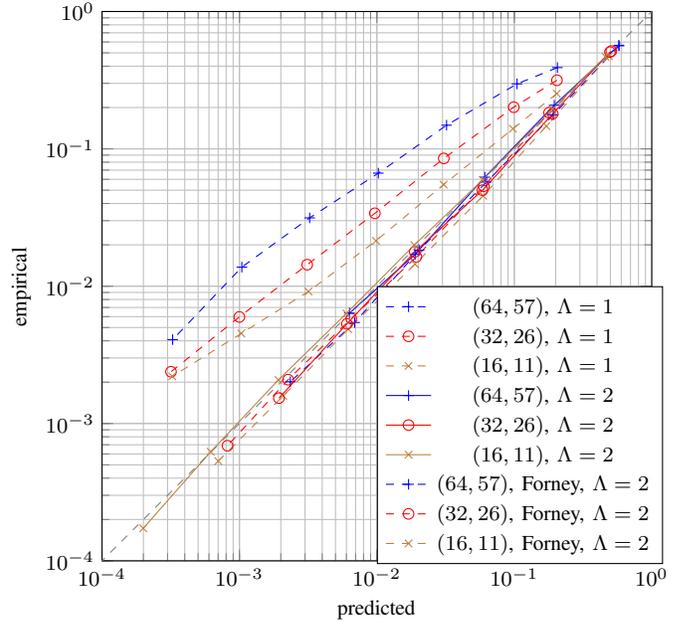

We begin by considering the quality of the blockwise SO from GCD-ORB for codes of different structure with binary modulation in an additive white Gaussian noise channel. For good-list sizes, $\Lambda\in\{1,2\}$, and RLCs of different $(n,k)$ dimensions, Fig. \ref{fig:blkSO_acc_listGCD_RLC} reports the conditional likelihood that the true codeword is not in the decoding list given the SO estimate that the decoding is not in the list having observed a large number of decodings. For all list sizes, the predicted probability of error accurately captures the empirical probability. Also shown is the approximation that would result from use of Forney's approximation, eq. \ref{eq:missingterm} and $\Lambda=2$, which is highly inaccurate.

\begin{figure}
\centering
	\footnotesize
	\begin{tikzpicture}[scale=1]
\footnotesize
\begin{loglogaxis}[
legend style={at={(0,1)},anchor= north west},
ymin=1e-4,
ymax=1,
width=3.5in,
height=3.5in,
grid=both,
xmin = 1e-4,
xmax = 1,
xlabel = {predicted},
ylabel = {empirical},
]

\addplot[blue,mark = +]
table[]{x y
0.57995 0.57936
0.1958 0.19355
0.060663 0.062208
0.019092 0.01935
0.006112 0.0055951
0.001987 0.00185616
};\addlegendentry{$(64,57)$, $\Lambda=4$}

\addplot[red,mark=o]
table[]{x y
0.507 0.50716
0.19162 0.19148
0.058818 0.054787
0.018645 0.017984
0.0059195 0.0059229
0.0018913 0.0017032
0.00060332 0.00062957
};\addlegendentry{$(32,26)$, $\Lambda=4$}

\addplot[brown,mark=x]
table[]{x y
0.47314 0.47718
0.18959 0.18924
0.058321 0.055632
0.018417 0.016979
0.0058541 0.0058792
0.0018656 0.0017549
0.00059354 0.00057203
0.00018906 0.00018148
};\addlegendentry{$(16,11)$, $\Lambda=4$}

\addplot[blue,mark = +,dotted,mark options=solid]
table[]{x y
0.58214 0.58019
0.19396 0.17607
0.060599 0.054834
0.019126 0.01705
0.0061228 0.0055147
0.0019441 0.0018416
};\addlegendentry{$(64,57)$, $\Lambda=2$}

\addplot[red,mark=o,dotted,mark options=solid]
table[]{x y
0.51183 0.51358
0.19456 0.18936
0.05776 0.059469
0.018947 0.018364
0.0060164 0.0055351
0.0018899 0.0016674
};\addlegendentry{$(32,26)$, $\Lambda=2$}

\addplot[brown,mark=x,dotted,mark options=solid]
table[]{x y
0.49153 0.45276
0.18632 0.16552
0.056988 0.045864
0.017976 0.013142
0.0058701 0.0044721
0.0018754 0.00149928
0.00060223 0.00046542
0.00018841 0.00014992
};\addlegendentry{$(16,11)$, $\Lambda=2$}

\addplot[blue,mark = +,dashed,mark options=solid]
table[]{x y
0.57318 0.56865
0.19451 0.13263
0.06217 0.015306
0.020214 0.0010616
};\addlegendentry{$(64,57)$, $\Lambda=1$}

\addplot[red,mark=o,dashed,mark options=solid]
table[]{x y
0.51452 0.51069
0.1803 0.1099
0.060305 0.012783
0.01948 0.00097895
};\addlegendentry{$(32,26)$, $\Lambda=1$}

\addplot[brown,mark=x,dashed,mark options=solid]
table[]{x y
0.49299 0.48574
0.17075 0.098321
0.05927 0.009048
0.018887 0.00073434
};\addlegendentry{$(16,11)$, $\Lambda=1$}

\addplot[gray,dashed]
table[]{x y
1e-5 1e-5
1 1
};

\end{loglogaxis}

\end{tikzpicture}
	\caption{SO Predicted BLER vs. empirical BLER: eBCH codes, SO-GCD-ORB, $E_b/N_0=3$.}
	\label{fig:blkSO_acc_listGCD_eBCH}
\end{figure}
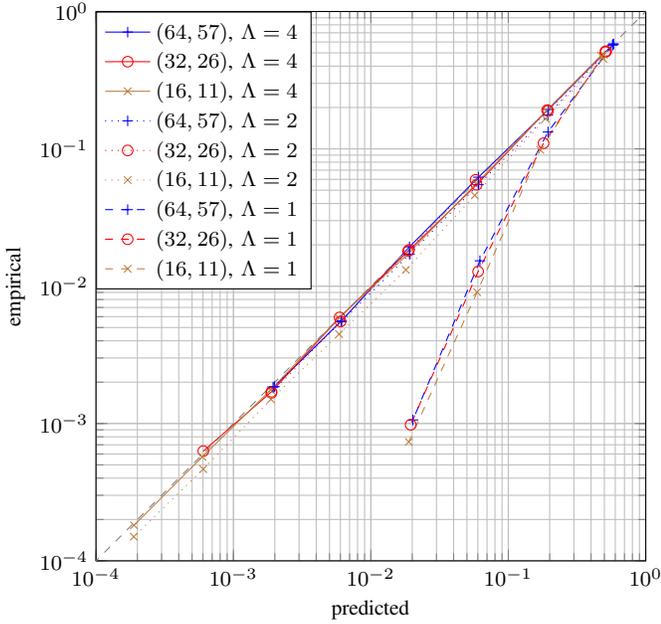

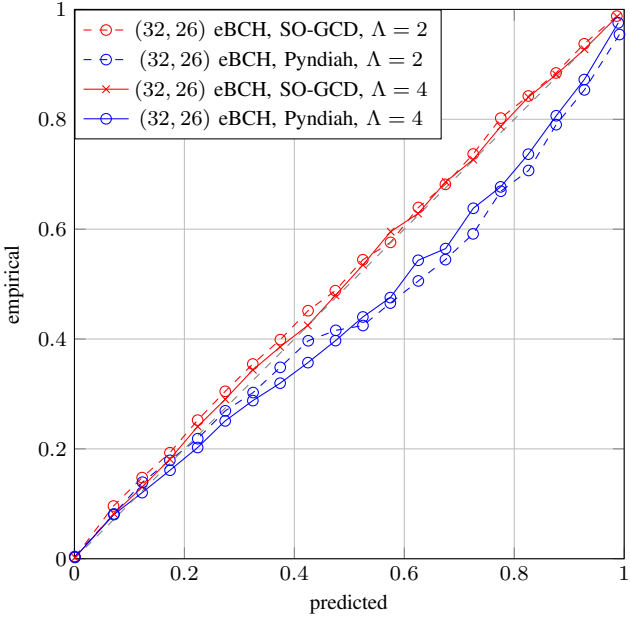
\begin{figure}
\centering
	\footnotesize
	\begin{tikzpicture}[scale=1]
\footnotesize
\begin{axis}[
legend style={at={(0,1)},anchor= north west},
legend columns=1,
ymin=0,
ymax=1,
width=3.5in,
height=3.5in,
grid=both,
xmin = 0,
xmax = 1,
xlabel = {predicted},
ylabel = {empirical},
]

\addplot[red,mark = o,dashed,mark options=solid]
table[]{x y
0.0011489 0.002862
0.071034 0.095974
0.12317 0.14771
0.17367 0.19307
0.22409 0.25234
0.27432 0.30484
0.3244 0.35452
0.37461 0.39891
0.42495 0.45158
0.47455 0.488
0.52463 0.54433
0.57493 0.57592
0.62568 0.63947
0.67479 0.68158
0.7253 0.73681
0.77547 0.80216
0.82564 0.84241
0.87606 0.88432
0.92733 0.93743
0.98659 0.98788
};\addlegendentry{$(32,26)$ eBCH, SO-GCD, $\Lambda=2$}

\addplot[blue,mark = o,dashed,mark options=solid]
table[]{x y
0.00039136 0.0039966
0.071888 0.081634
0.12377 0.13935
0.17357 0.17919
0.22426 0.21856
0.2745 0.26935
0.32498 0.30251
0.37431 0.34845
0.42528 0.3968
0.47537 0.41576
0.52485 0.4246
0.57465 0.46539
0.62555 0.50573
0.67484 0.54451
0.72516 0.59144
0.77564 0.66903
0.82613 0.70679
0.87623 0.79025
0.92762 0.85351
0.9914 0.95431
};\addlegendentry{$(32,26)$ eBCH, Pyndiah, $\Lambda=2$}

\addplot[red,mark = x]
table[]{x y
0.0011415 0.0021609
0.071657 0.081821
0.12291 0.13035
0.17353 0.18113
0.22412 0.24073
0.27428 0.2903
0.32456 0.34386
0.37408 0.38557
0.42448 0.42483
0.47468 0.47946
0.52477 0.53494
0.57523 0.59554
0.62502 0.62839
0.67529 0.68606
0.72526 0.72669
0.77549 0.78701
0.82579 0.84033
0.87641 0.88144
0.92771 0.92746
0.98714 0.987
};\addlegendentry{$(32,26)$ eBCH, SO-GCD, $\Lambda=4$}

\addplot[blue,mark = o]
table[]{x y
0.00084149 0.0023519
0.071682 0.080302
0.12311 0.12037
0.17387 0.16113
0.22414 0.20246
0.27416 0.25088
0.32453 0.28789
0.37417 0.31958
0.42473 0.35703
0.4749 0.39704
0.52467 0.44016
0.57459 0.47543
0.62524 0.54333
0.6751 0.56457
0.7255 0.63799
0.77531 0.67698
0.82566 0.73654
0.87642 0.80647
0.92742 0.87234
0.98922 0.97555
};\addlegendentry{$(32,26)$ eBCH, Pyndiah, $\Lambda=4$}

\addplot[gray,dashed]
table[]{x y
0 0
1 1
};

\end{axis}

\end{tikzpicture}
	\caption{SO Predicted BER vs. empirical BER: eBCH codes, SO-GCD-ORB, $E_b/N_0=3$.}
	\label{fig:bitSO_acc_listGCD}
\end{figure}

Fig. \ref{fig:blkSO_acc_listGCD_eBCH} shows equivalent results for high-rate eBCH codes that are often used in product code constructions \cite{pyndiah_1998,condo2022_iterative,galligan2023_block}. For a single decoding, the estimator is pessimistic, owing to the structured deviance from a random code, but larger lists are typically necessary for product code decoding. Once the good-list size, $\Lambda$, is $2$ or larger, the deviance disappears and the estimate is notably accurate. The precision of the blockwise SO translates to accurate bitwise SO, as illustrated in Fig. \ref{fig:bitSO_acc_listGCD}, with the enhancement over the traditional Pynidah approximation being evident.

\section{SISO Decoding}
\label{sec:peva}
An efficient way of making long, powerful SI error correction codes is to concatenate shorter component codes for which there is an accurate SISO decoder. Elias's product codes \cite{elias_error-free_1954} are one particularly appealing approach, as both their encoding and decoding can be highly parallelized, resulting in low latency. With an $(n,k)$ component code, $k^2$ information bits are written into a $k\times k$ square matrix, each of the $k$ rows are encoded, and each of the resulting $n$ columns are encoded, resulting in an $(n^2,k^2)$ code. 

In 1998, Pyndiah proposed iterative SISO decoding of product codes, also known as turbo product codes, \cite{pyndiah_1998}. For the original decoder, the component code was required to have an efficient hard detection decoder resulting in a restricted class. Recently, the performance of product codes with polar component codes have been considered with successive cancellation list decoding using Pyndiah's soft update \cite{bioglio2019construction,condo2020practical,cocskun2024precoded} or GRAND-like SO \cite{yuan2024nearoptimal}, which gives better performance. SI GRAND algorithms have also been applied with Pyndiah's soft update solely as true list decoders \cite{condo2022_iterative, galligan2023_block} or, more recently, with GRAND SO \cite{yuan2023soft}, which results in significantly enhanced performance. As GRAND algorithms can decode any moderate redundancy component code, they open up a broad palette of product code constructions.

Here we consider block turbo decoding with SO-GCD-ORB. As with GRAND, GCD can decode any linear component code and, armed with accurate SO from considering it as a GRAND variant, it can be used for iterative decoding. Block turbo decoding of product codes works as follows:
\begin{itemize}
    \item[0] The channel LLRs are stored in an $n\times n$ matrix $\mathbf{L}_\text{Ch}$. The a priori LLRs of the coded bits are initialized to zero, i.e., $\mathbf{L}_\text{A}=\mathbf{0}$.
    \item[1] Each row of $\mathbf{L}_\text{Ch}+\mathbf{L}_\text{A}$ is processed by a \ac{SISO} decoder, and the resulting APP LLRs and extrinsic LLRs are stored in the corresponding rows of $\mathbf{L}_\text{APP}$ and $\mathbf{L}_\text{E}$, respectively. A hard decision is made based on $\mathbf{L}_\text{APP}$. If all rows and columns of the decision correspond to valid codewords, the block turbo decoder returns the hard output, indicating successful decoding. Otherwise, $\mathbf{L}_\text{A}$ is set to $\alpha\mathbf{L}_\text{E}$, for some $\alpha>0$, and the decoder proceeds to the column update. 
    \item[2] Each column of $\mathbf{L}_\text{Ch}+\mathbf{L}_\text{A}$ is decoded and the columns of $\mathbf{L}_\text{APP}$ and $\mathbf{L}_\text{E}$ are updated as in step $1$. A hard decision is performed using $\mathbf{L}_\text{APP}$. If the obtained binary matrix is valid, decoding success is declared. If the maximum iteration count is reached, a decoding failure is returned. Otherwise, we set $\mathbf{L}_\text{A} = \alpha\mathbf{L}_\text{E}$ and proceed to the next iteration (i.e., return to step 1).
\end{itemize}
In simulation, we set $\alpha=0.5$,  and use SO-GCD-ORB as the SISO decoder with eq. \eqref{eq:LLRi}. Note that, in practice, each row is decodable in parallel and each column is decodable in parallel, resulting in desirable low-latency.

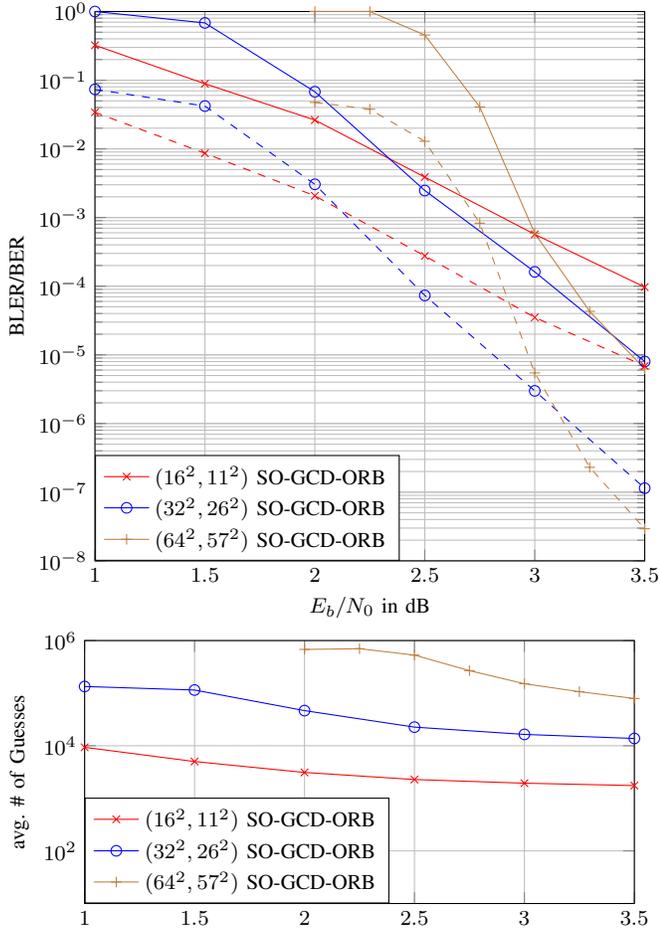
\begin{figure}[t]
	\centering
	\begin{tikzpicture}[scale=1]
\footnotesize
\begin{semilogyaxis}[
legend style={at={(0,0)},anchor= south west},
legend columns=1,
ymin=1e-8,
ymax=1,
width=3.5in,
height=3.5in,
grid=both,
xmin = 1,
xmax = 3.5,
xlabel = $E_b/N_0$ in dB,
ylabel = {BLER/BER},
]

\addplot[red,mark=x]
table[]{x y
0 0.9375
0.5 0.66667
1 0.32258
1.5 0.088757
2 0.02627
2.5 0.0038765
3 0.00056871
3.5 9.7087e-05
4 1.1511e-05
};\addlegendentry{$(16^2,11^2)$ SO-GCD-ORB}

\addplot[blue,mark=o]
table[]{x y
1 1
1.5 0.68182
2 0.067873
2.5 0.002474
3 0.00016164
3.5 8e-06
};\addlegendentry{$(32^2,26^2)$ SO-GCD-ORB}

\addplot[brown,mark=+]
table[]{x y
2 1
2.25 1
2.5 0.45455
2.75 0.040706
3 0.0005877
3.25 4.2981e-05
3.5 6.2399e-06
};\addlegendentry{$(64^2,57^2)$ SO-GCD-ORB}

\addplot[red,mark=x,dashed,mark options=solid]
table[]{x y
0 0.1015
0.5 0.068136
1 0.034035
1.5 0.0086312
2 0.0020697
2.5 0.00027552
3 3.5094e-05
3.5 6.8736e-06
4 7.8641e-07
};

\addplot[blue,mark=o,dashed,mark options=solid]
table[]{x y
1 0.07357
1.5 0.041992
2 0.0030456
2.5 7.3318e-05
3 2.9889e-06
3.5 1.1432e-07
};

\addplot[brown,mark=+,dashed,mark options=solid]
table[]{x y
2 0.047471
2.25 0.037765
2.5 0.012876
2.75 0.00082731
3 5.4581e-06
3.25 2.3032e-07
3.5 2.9311e-08
};

\end{semilogyaxis}
\end{tikzpicture}

\begin{tikzpicture}[scale=1]
\footnotesize
\begin{semilogyaxis}[
legend style={at={(0,0)},anchor= south west},
legend columns=1,
ymin=1e1,
ymax=1e6,
width=3.5in,
height=2in,
grid=both,
xmin = 1,
xmax = 3.5,
ylabel = {avg. \# of Guesses},
]

\addplot[red,mark=x]
table[]{x y
0 19629
0.5 15082
1 9333.9
1.5 5006
2 3106.6
2.5 2277.9
3 1940.6
3.5 1749.1
4 1607.4
};\addlegendentry{$(16^2,11^2)$ SO-GCD-ORB}

\addplot[blue,mark=o]
table[]{x y
1 1.3384e+05
1.5 1.1465e+05
2 46545
2.5 22595
3 16422
3.5 13746
};\addlegendentry{$(32^2,26^2)$ SO-GCD-ORB}

\addplot[brown,mark=+]
table[]{x y
2 6.7827e+05
2.25 6.9937e+05
2.5 5.2934e+05
2.75 2.6781e+05
3 1.5109e+05
3.25 1.0666e+05
3.5 7.9120e+04
};\addlegendentry{$(64^2,57^2)$ SO-GCD-ORB}

\end{semilogyaxis}

\end{tikzpicture}
	\caption{Turbo product code decoding with SO-GCD-ORB. For eBCH product codes of dimensions $(256,121)$, $(1024,676)$ and $(4096,3249)$, the upper panel shows block (solid) and bit (dashed) error rate as well as achievability bounds. The lower panel shows the average number of guesses until the decoding terminates.}
	\label{fig:product_GCD}
\end{figure}
Fig. \ref{fig:product_GCD} shows performance of powerful eBCH product codes of dimensions $(256,121)$, $(1024,676)$ and $(4096,3249)$, constructed using component codes from Section \ref{sec:SOGCD}. The upper panel shows block- and bit-error-rate curves, which offer comparable performance to SOGRAND \cite{yuan2023soft}. The lower panel shows the average number of noise effect queries until a decoding is found with SO-GCD-ORB, which indicates that low latency, low-energy implementations following the 1-line OBRGRAND query order would be possible in hardware \cite{Riaz23}. 

\section{Discussion}
SI error correction decoders that can provide accurate blockwise or bitwise SO provide significant extra functionality. Blockwise SO can be used to accurately trigger ARQ requests or flag data for erasure correction. Bitwise SO can be used to create SISO decoders that can be used to efficiently decode long, high redundancy codes. Few decoders provide both. Recently, it has been shown that accurate blockwise and bitwise can be efficiently evaluated during a SI GRAND decoding \cite{galligan2023upgrade,yuan2023soft} and the methodology has been extended to successive cancellation decoders \cite{yuan2024nearoptimal}. By considering GCD as a sophisticated version of GRAND that makes parallelized queries, we establish that SO can be produced with it too. The accuracy of that SO is confirmed as well as a demonstration of the powerful applications that result from its use.

\bibliographystyle{IEEEtran}
\bibliography{references}

\end{document}